\documentclass[aps,prl,amsmath,amssymb,superscriptaddress,preprintnumbers,twocolumn,
 amsmath,amssymb,
 aps,
]{revtex4-2}

\usepackage{graphicx}
\usepackage{dcolumn}
\usepackage{bm}
\usepackage{cancel}

\usepackage[usenames, dvipsnames]{xcolor}
\usepackage{tikz}
\usetikzlibrary{shapes}
\usepackage[co mpat=1.1.0]{tikz-feynman}

\usepackage{hyperref}

\begin{document}

\preprint{}

\title{Asymptotic Ultraviolet-safe Unification of Gauge and Yukawa Couplings: \\
The exceptional case}

\author{Giacomo Cacciapaglia}
\email{g.cacciapaglia@ipnl.in2p3.fr}
\affiliation{Université de Lyon, Université Claude Bernard Lyon 1, CNRS/IN2P3, IP2I Lyon, UMR 5822, Villeurbanne F-69622, France}

\author{Aldo Deandrea}
\email{deandrea@ip2i.in2p3.fr}
\affiliation{Université de Lyon, Université Claude Bernard Lyon 1, CNRS/IN2P3, IP2I Lyon, UMR 5822, Villeurbanne F-69622, France}
\affiliation{Department of Physics, University of Johannesburg, PO Box 524, Auckland Park 2006, South Africa}

\author{Roman Pasechnik}
\email{roman.pasechnik@hep.lu.se}
\affiliation{Department of Physics, Lund
University, S\"olvegatan 14A, SE-223 62 Lund, Sweden}

\author{Zhi-Wei Wang}
\email{zhiwei.wang@uestc.edu.cn}
\affiliation{School of Physics, University of Electronic Science and Technology of China, 88 Tian-run Road, Chengdu, China}

\begin{abstract}
The ultimate dream of unification models consists in combining both gauge and Yukawa couplings into one unified coupling. This is achieved by using a supersymmetric exceptional ${\rm E}_6$ gauge symmetry together with asymptotic unification in compact five-dimensional space-time. The  ultraviolet fixed point requires \emph{exactly three} fermion generations: one in the bulk, and the two light ones localised on the ${\rm SO}(10)$ boundary in order to cancel gauge anomalies. A second option allows to preserve baryon number and to lower the compactification scale down to the typical scales of the intermediate Pati-Salam gauge theory.
\end{abstract}

\keywords{Ultraviolet fixed point; ${\rm E}_6$ grand unification; Pati-Salam theory}
\maketitle

Coupling constants quantify the strength of interactions among particles and evolve with the energy of the collisions via the renormalisation group \cite{Wilson:1971bg,Wilson:1971dh}.
The term unification usually indicates the crossing of the gauge couplings at a specific energy scale, assuming that a new larger gauge symmetry emerges after that point. This mechanism yields a unified description of the electroweak and strong interactions as a single force, within the so-called Grand Unified Theories (GUTs) \cite{Georgi:1974sy,Georgi:1974yf,Amaldi:1991zx}. GUTs feature the assembling of matter as well as gauge degrees of freedom within multiplets in a rather elegant and dangerous way. In fact, gauge unification typically implies relations among the Standard Model (SM) couplings and the presence of new gauge states, often bringing embarrassing predictions, such as fast proton decay \cite{Langacker:1980js}. While countermeasures to these shortcomings are well-known \cite{Langacker:1980js,Raby:2017ucc}, they typically tend to partially obscure the beauty and the simplicity of the original models: the unification scale is pushed close to the Planck mass and fields in large representations of the unified group are often a must \cite{Aulakh:2003kg}. The latter leads to an unresolved issue: GUTs, especially based on supersymmetry, feature Landau poles for the unified coupling evolution soon after unification, hence maiming the validity of the theory.

The idea of asymptotic unification solves this problem \cite{Bajc:2016efj}, as it is based on the requirement that the theory should flow to a non-trivial ultraviolet (UV) fixed point at high energy, without the need of a specific unification scale. This is possible in models with a single extra space dimension \cite{Gies:2003ic,Morris:2004mg}, which yields a power running of the gauge couplings \cite{Dienes:1998vh,Hebecker:2004xx}. The extra dimension is compactified on an interval, where the bulk could be flat or warped \cite{Randall:1999ee} as the high-energy behaviour is the same. The model features a five-dimensional unified gauge symmetry in the bulk, so that at high energies (small distances) the unified behaviour emerges. The bulk gauge group is broken by boundary conditions, and the SM fields are identified with the zero modes of the bulk fields, implying a completely different arrangement of matter and gauge fields within multiplets as compared to traditional unification. 

Extra-dimensional GUTs were widely explored \cite{Kawamura:1999nj,Kawamura:2000ir,Hall:2001xb,Hebecker:2001wq,Kim:2002im}, where the unification is due to a sub--leading logarithmic running of the gauge couplings \cite{Contino:2001si}. Asymptotic unification, instead, is driven by the contributions of complete multiplets, not by incomplete ones as in the standard GUT. Hence, it is the contribution of the bulk resonances that drives it \cite{Dienes:1998vh,Dienes:2002bg}. Asymptotic unification is a more natural choice than the standard one in extra dimensional models: in fact, at energies above the inverse radius of the compact dimension, the theory approaches the extra-dimensional evolution, showing that the electroweak and strong interactions have always been a single force at small distances. They appear different at low energies only due to the breaking effect of the compactification. A concrete model of asymptotic Grand Unification (aGUT) was recently proposed, based on the minimal $\text{SU}(5)$ gauge group \cite{Cacciapaglia:2020qky,Cacciapaglia:2022nwt}\footnote{Note that alternative asymptotically safe GUT models may be realised \cite{Molinaro:2018kjz,Wang:2018yer,Sannino:2019sch} via large $N_f$ methods \cite{Antipin:2017ebo,Mann:2017wzh}.}. Interestingly, the running of the Yukawa couplings imposes critical conditions on these models, for instance ruling out the traditional $\text{SO}(10)$ symmetry \cite{Khojali:2022gcq}.

A more ambitious programme would require the unification of all the couplings of the SM, namely gauge and Yukawa couplings, at least for one generation. In extra dimensions, this has been explored in the context of gauge-Higgs unification models \cite{Hosotani:2015hoa}. Within the asymptotic unification paradigm, this can be achieved if the UV fixed points for all the couplings coincide. Supersymmetry comes in handy, as it naturally relates couplings of bosons (gauge bosons) to those of fermions (gauginos). Within this framework, the Lie group ${\rm E}_6$ \cite{Gursey:1975ki,Babu:2015psa} offers an exceptional opportunity, as the SM fermions can be embedded within the adjoint representation, hence stemming from ${\rm E}_6$ gauginos \cite{Kobayashi:2004ya}. As the Yukawa couplings are generated from the ${\rm E}_6$ gauge interactions, they share the same fixed point in the UV.

\begin{figure}[t!]
\centering
\includegraphics[width=6.0cm]{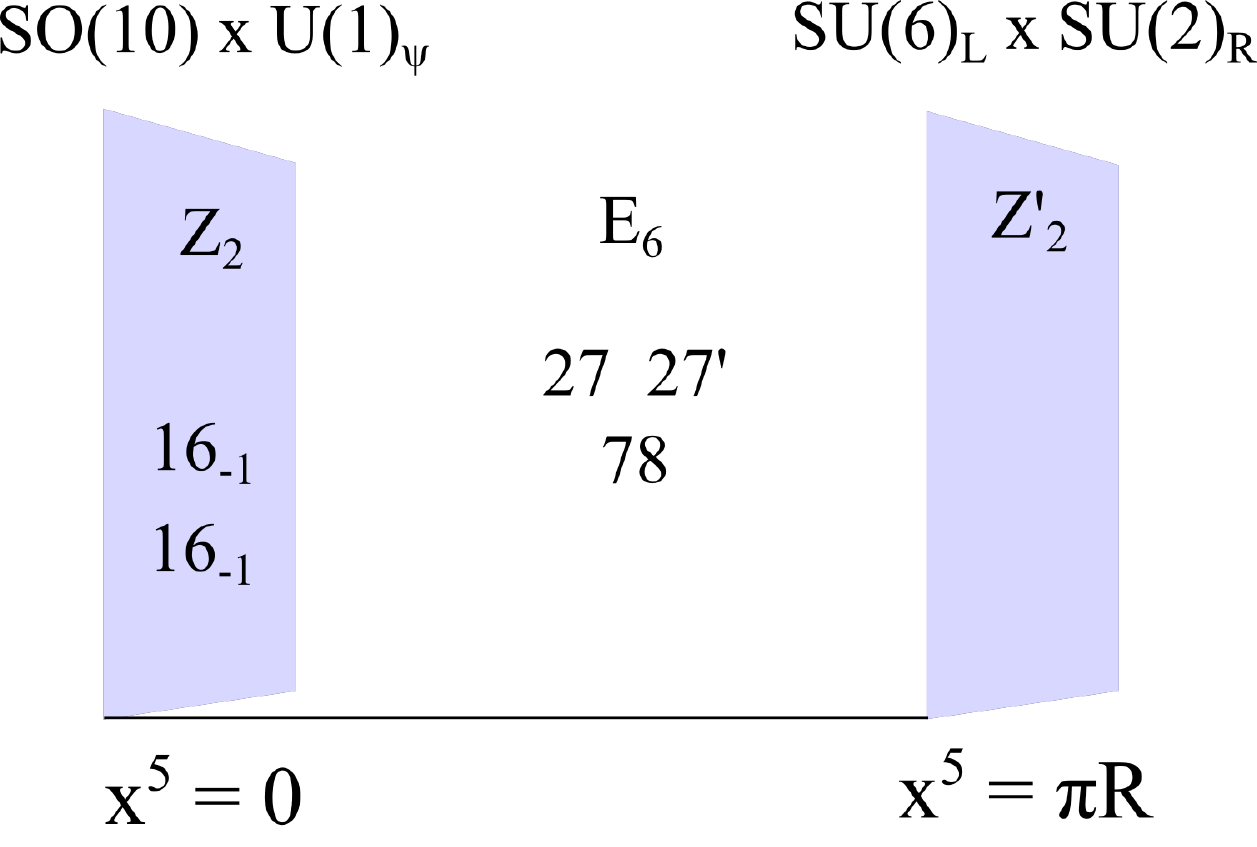}
\caption{\label{fig:cartoon} Illustration of the ${\rm E}_6$ aGUT  with 5D ${\rm E}_6$ hypermultiplets in the bulk and ${\rm SO}(10)$ $\mathcal{N}=1$ supermultiplets on the $x^5=0$ boundary, as required by gauge anomaly cancellation.}
\end{figure}

\begin{figure*}[t!]
\centering
\includegraphics[width=17cm]{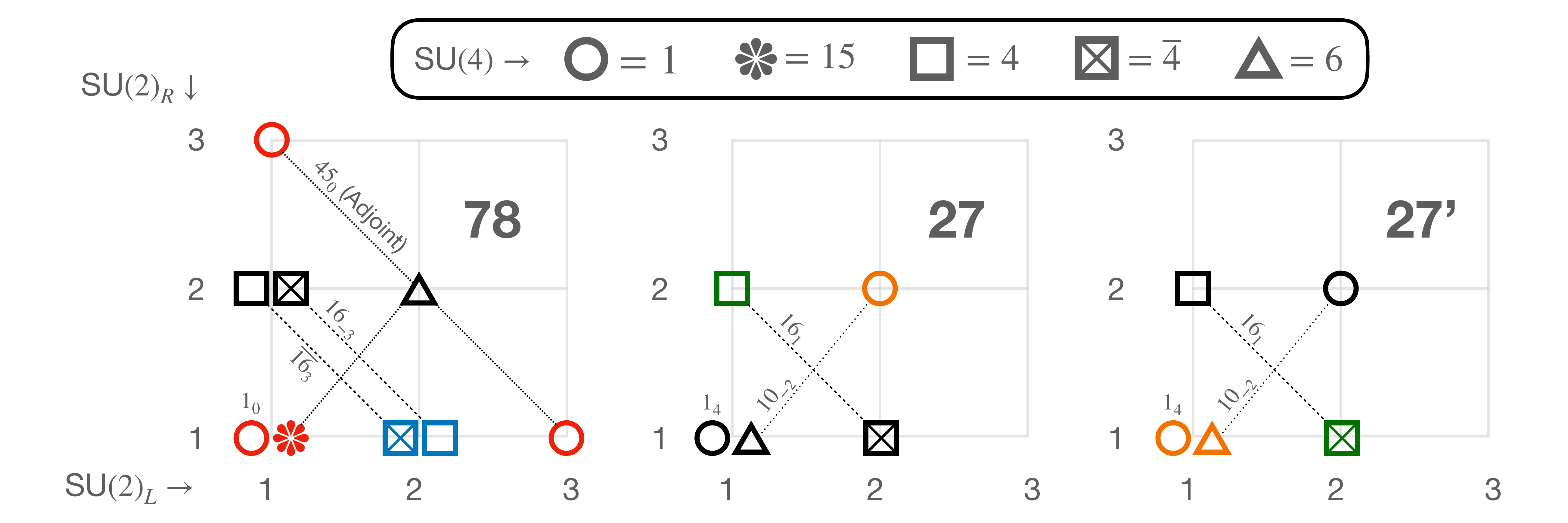}
\caption{\label{fig:irreps} Illustration of the three bulk fields in terms of PS$\times\text{U}_\psi$ components of the ${\rm E}_6$ representations. The position on the grid indicates the quantum numbers under $\text{SU}(2)_\text{L}\times\text{SU}(2)_\text{R}$, while the symbols represent $\text{SU}(4)$ representations. The $\text{SO}(10)$ components are linked by dashed lines and labelled. Finally, the colours indicate the presence of a zero mode: red for the $W^\alpha_{78}$ component and blue for the $\Phi_{78}$, green for the matter $\Phi$ and orange for the $\Phi^c$. The black symbols correspond to components without a zero mode.}
\end{figure*}

The model, therefore, consists of a supersymmetric ${\rm E}_6$ gauge theory in five dimensions (5D) with matter fields in the fundamental ${\bf 27}$. The fifth dimension is compactified on the orbifold $S^1/\mathbb{Z}_2\times\mathbb{Z}_2'$. From the four-dimensional (4D) point of view, this theory has $\mathcal{N}=2$ supersymmetry \cite{Mirabelli:1997aj}, hence the gauge and matter superfields consist of
\begin{eqnarray}
    \mbox{Gauge :} & \Rightarrow & W^\alpha_{78} + \Phi_{78}\,, \\
    \mbox{Matter :} & \Rightarrow & \Phi_{27} + \Phi^c_{27}\,,
\end{eqnarray}
where $W^\alpha$ is the vector superfield, while $\Phi$ are the chiral superfields, with $\Phi^c_{27}$ having conjugate quantum numbers as compared to $\Phi_{27}$. The model Lagrangian can be constructed in terms of these components \cite{Mirabelli:1997aj,Arkani-Hamed:2001vvu,Hebecker:2001ke}. The orbifold projection, encoded in the two parities $\mathbb{Z}_2$ and $\mathbb{Z}_2'$ centred on the two endpoints of the interval, break both the $\mathcal{N}=2$ supersymmetry to $\mathcal{N}=1$ in 4D and the gauge symmetry. For the latter, there are three possible patterns \cite{Braam:2010sy}:
\begin{eqnarray}
    \mbox{A:} &\quad& \text{E}_6 \to \text{SO}(10) \times \text{U}(1)_\psi\,,  \nonumber \\
     \mbox{B:} &\quad& \text{E}_6 \to \text{SU}(6)_\text{L} \times \text{SU}(2)_\text{R}\,,  \\
     \mbox{C:} &\quad& \text{E}_6 \to \text{SU}(6)_\text{R} \times \text{SU}(2)_\text{L}\,,  \nonumber
\end{eqnarray}
where the subscript ``L'' and ``R'' refer to the custodial symmetry in the SM. Choosing any pair of the above patterns for the two orbifold projections leads to a 4D gauge theory based on Pati-Salam (PS) \cite{Pati:1974yy} times an additional abelian $\text{U}(1)$ symmetry: $\text{SU}(4) \times \text{SU}(2)_\text{L} \times \text{SU}(2)_\text{R} \times \text{U}(1)_\psi$. However, only one choice is phenomenologically viable. For the combination B-C, it is not possible to obtain chiral zero modes for the SM fermions, while for A-C the zero mode spectrum does not allow for the breaking of the PS symmetry. Hence, the only viable model is based on A-B as illustrated in Fig.~\ref{fig:cartoon}. The field content is summarised in Fig.~\ref{fig:irreps}, where we highlight the decomposition with respect to the various subgroups of ${\rm E}_6$, and we include two matter fields in the fundamental representation, ${\bf 27}$ and ${\bf 27'}$, with different orbifold parities. As we will see below this is the minimal and unique embedding of one SM generation in the bulk.

A similar set-up was first proposed as a string-inspired standard GUT \cite{Kobayashi:2004ya}. Our proposal differs in two crucial features: a) the unification is driven asymptotically by the UV fixed point; b) all non-SM zero mode fields receive mass without the need for additional bulk fields. The second feature is crucial to maintain the UV fixed point of the theory.

The UV fixed point emerges from the power law corrections to the renormalisation group evolution of the effective 4D gauge coupling \cite{Dienes:1998vh}. The beta function can be expressed as an effective 5D 't Hooft coupling \cite{Dienes:2002bg}, defined in terms of the 4D coupling $\alpha = g^2/4\pi$ as
\begin{equation} \label{eq:5Dcoup}
 \tilde{\alpha} (\mu) = \alpha (\mu)\ \frac{\mu}{m_\text{KK}}\,,   
\end{equation}
where $\mu$ is the renormalisation scale and $m_\text{KK}$ is the mass of the first Kaluza-Klein (KK) state. At one loop, the beta function for $\tilde{\alpha}$ reads
\begin{equation}
    \frac{d \tilde{\alpha}}{d \ln \mu} = \tilde{\alpha} - \frac{b_5}{2 \pi} \tilde{\alpha}^2\,,
\end{equation}
which is valid for $\mu \gg m_\text{KK}$ and has a UV zero at $\tilde{\alpha}^\ast_\text{UV} = 2 \pi/b_5$ for $b_5 > 0$. For our model, we find \cite{Flacke:2003ac}
\begin{equation} \label{eq:RGE5D}
    b_5 = \frac{\pi}{2} \left(C(G) - \sum_i T_i (R_i) \right) = 3 \pi\,,
\end{equation}
where $C(G) = 12$ and $T(27) = 3$ for ${\rm E}_6$~\footnote{This result stems from a direct 5D loop computation. More commonly, $b_5$ is computed in terms of KK states, leading to $$b_5 = 2 \left( C(G) - \sum_i T(R_i) \right) = 12\,.$$ The difference is mainly due to the phase space integration in 4D versus 5D, and it can be considered as an uncertainty on the numerical predictions.}. Hence, only one more fundamental can be added in the bulk before the UV fixed point disappears. As a consequence, the model in Ref.\cite{Kobayashi:2004ya} does not have such UV fixed point. In our model, instead, the UV fixed point remains perturbative, hence endorsing the model with enhanced calculability. We remark that this behaviour remains the same for both flat and warped extra dimensions~\footnote{The two geometries have the same short distance behaviour. Hence, the only difference will be encoded in threshold effects near the compactification scale.}.

From the 4D point of view, the theory is invariant under PS$\times \text{U}(1)_\psi$, which we use to classify the relevant states. At the zero mode level, besides the gauge superfields in $W^\alpha_{78}$, the chiral superfields contain the following states:
\begin{eqnarray}
    \Phi_{78} &\supset& (\bar{\bf 4},{\bf 1},{\bf 2})_{-3}  + ({\bf 4},{\bf 1},{\bf 2})_{3}\,, \\
    \Phi_{27} &\supset& ({\bf 4},{\bf 2},{\bf 1})_{1}\,,  \\
    \Phi^c_{27} &\supset& ({\bf 1},{\bf 2},{\bf 2})_{2}\,,  \\
    \Phi_{27'} &\supset& (\bar{\bf 4},{\bf 1},{\bf 2})_{1}\,, \\
    \Phi_{27'}^c &\supset& ({\bf 6},{\bf 1},{\bf 1})_{2}\, +\, ({\bf 1},{\bf 1},{\bf 1})_{-4} \,,
\end{eqnarray}
shown as coloured symbols in Fig.~\ref{fig:irreps}. We can already see that the ${\bf 27}$ contains the left-handed SM fermions $({\bf 4},{\bf 2},{\bf 1})_{1}$ and two Higgs doublets in $({\bf 1},{\bf 2},{\bf 2})_{2}$. The PS Yukawa couplings stem from the gauge interactions of the ${\bf 27}$ with the $\Phi_{78}$ component of the gauge supermultiplet \cite{Anderson:1999em,Anderson:2000ni}, leading to:
\begin{equation} \label{eq:gauge27}
    g\ \Phi_{27}^c \Phi_{78} \Phi_{27} \supset \frac{g}{\sqrt{2}} ({\bf 1},{\bf 2},{\bf 2})_{2}\ (\bar{\bf 4},{\bf 1},{\bf 2})_{-3}\ ({\bf 4},{\bf 2},{\bf 1})_{1}\,,
\end{equation}   
thus the right-handed SM fermions must be the $(\bar{\bf 4},{\bf 1},{\bf 2})_{-3}$ component of the gauge $\Phi_{78}$. The gauge interactions of the ${\bf 27'}$ contain the following zero-mode terms:
\begin{multline} \label{eq:gauge27X}
    g\ \Phi_{27'}^c \Phi_{78} \Phi_{27'} \supset - \frac{g}{\sqrt{2}} ({\bf 1},{\bf 1},{\bf 1})_{-4}\ ({\bf 4},{\bf 1},{\bf 2})_{3}\ (\bar{\bf 4},{\bf 1},{\bf 2})_{1} \\
    + \frac{g}{\sqrt{2}} ({\bf 6},{\bf 1},{\bf 1})_{2}\ (\bar{\bf 4},{\bf 1},{\bf 2})_{-3}\ (\bar{\bf 4},{\bf 1},{\bf 2})_{1}\,.
\end{multline}
The importance of the couplings above is related to the breaking of the PS$\times \text{U}(1)_{\psi}$ gauge group down to the SM one. In fact, the $({\bf 1},{\bf 1},{\bf 1})_{-4}$ from the $\bf 27'$ can effectively break $\text{U}(1)_\psi$ and also give a mass to the  $({\bf 4},{\bf 1},{\bf 2})_3$ gauginos. Moreover, the presence of a $({\bf 4},{\bf 1},{\bf 2})_{3}$ state in the gauge multiplet allows for the breaking of the PS symmetry via the Scherk-Schwarz mechanism \cite{Scherk:1979zr}. The breaking of the gauge symmetry down to the SM, therefore, does not require additional states in this theory. The zero modes in the $({\bf 6},{\bf 1},{\bf 1})_{2}$ component, which behave like a vector-like bottom quark singlet, can only receive a mass via a superpotential localised on the $\text{SO}(10)$ boundary with parity A~\footnote{In terms of $\text{SO}(10)\times \text{U}(1)_\psi$ components, the superpotential term reads:
$$\left. \Phi_{27'}^c\right|_{10_2} \left. \Phi_{27'}^c\right|_{10_2} \left. \Phi_{27'}^c\right|_{1_{-4}}\,.$$}.

\begin{figure}[t!]
\centering
\includegraphics[width=8.5cm]{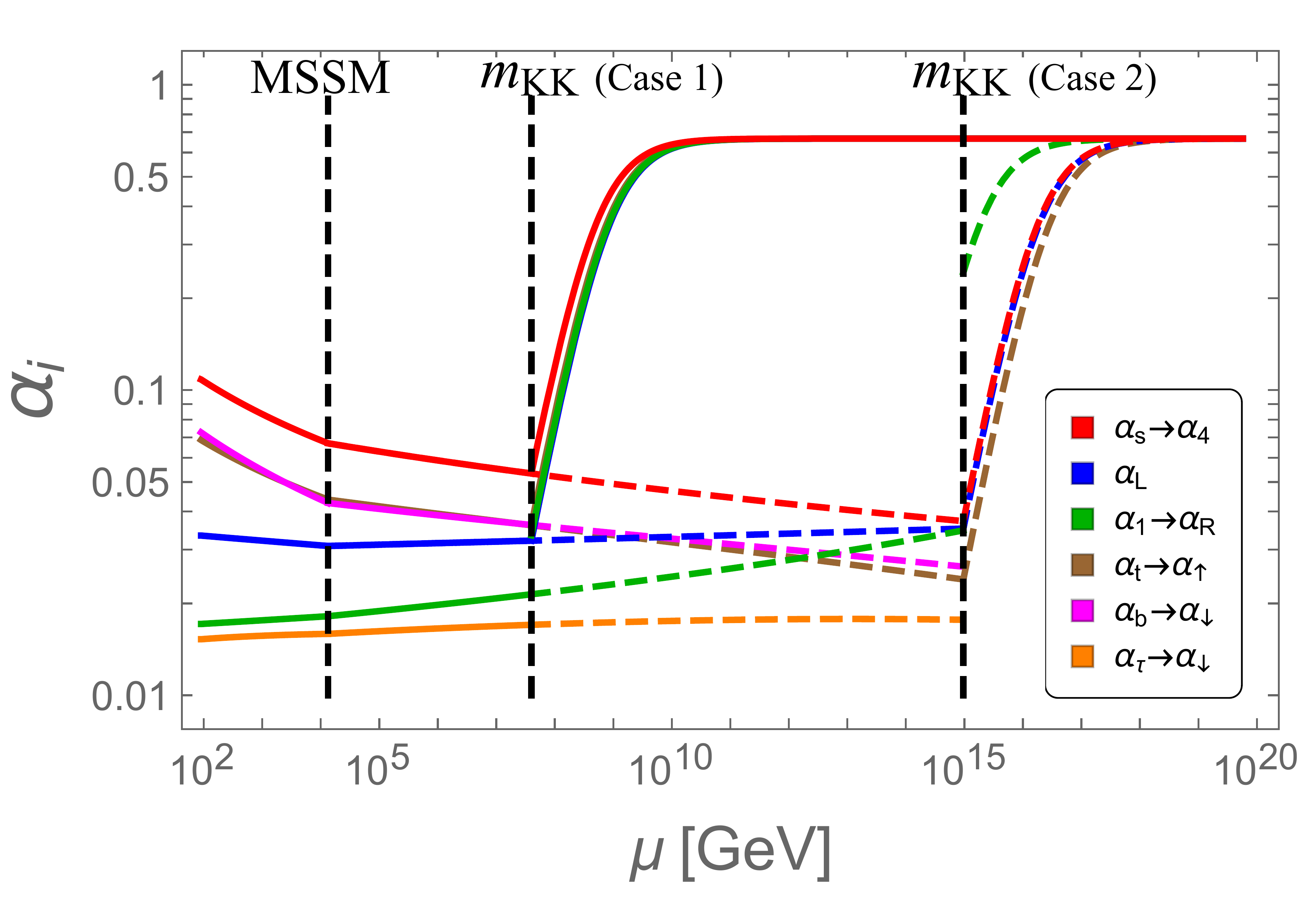}
\caption{\label{fig:running} Schematic running of the SM gauge couplings and third generation Yukawa couplings for two values of $m_\text{KK}$, where the Yukawa values correspond to $\tan \beta = 40$. The couplings are rescaled to match the ${\rm E}_6$ unification: $\alpha_x = 2 \frac{y_x^2}{4\pi}$ for Yukawas and $\alpha_1 = \frac{5}{3} \alpha'$ for hypercharge, where $\alpha_\text{R}$ comes from the usual PS matching.}
\end{figure}

At the UV fixed point, the matching between ${\rm E}_6$ and the PS$\times\text{U}(1)_\psi$ couplings reads:
\begin{equation} \label{eq:E6gauge}
g_4 = g_L = g_R \equiv g\,, \quad g_\psi = \frac{g}{\sqrt{2}}\,.
\end{equation}
For the SM Yukawa couplings in Eq.~\eqref{eq:gauge27}, we have
\begin{equation} \label{eq:E6yuk}
\quad y_{\uparrow} = y_{\downarrow} \equiv \frac{g}{\sqrt{2}}\,,  
\end{equation}
where $y_{\uparrow} = y_\text{t} = y_{\nu_\tau}$ is the Yukawa of up-type fermions, while $y_{\downarrow} = y_\text{b} = y_\tau$ for down-type ones. The identification of up and down-type Yukawas occurs at the PS-breaking scale, which is typically close to $m_\text{KK}$, while the relation between top and bottom mass also depends on the ratio of Higgs vacuum expectation values, as typical in supersymmetric models~\footnote{We recall that
$$y_\text{t} = \frac{m_\text{t}}{v \cot \beta}\,, \quad y_{\text{b},\tau} = \frac{m_{\text{b},\tau}}{v \sin \beta}\,,$$ where $v=246$~GeV is the Higgs vacuum expectation value in the SM.}. This ratio, expressed in terms of $\tan \beta$, requires
\begin{equation}
    \tan \beta = \frac{\langle H_d \rangle}{\langle H_u \rangle} = \frac{m_\text{t} (m_\text{KK})}{m_\text{b} (m_\text{KK})} \sim 40\,,
\end{equation}
where the masses are evaluated at the KK scale. In Fig.~\ref{fig:running} we show a schematic plot of the renormalisation group evolution of the SM couplings in the ${\rm E}_6$ model. For simplicity, we identify the PS and $\text{U}(1)_\psi$ breaking scales to $m_\text{KK}$, and fix the supersymmetry breaking scale to $10$~TeV, above which the minimal supersymmetric SM (MSSM) is a good description. The couplings correspond to the usual 4D ones up to the scale $m_\text{KK}$, above which they are replaced by the corresponding 5D 't Hooft couplings, defined in Eq.~\eqref{eq:5Dcoup}. Also, we plot the couplings rescaled to the ${\rm E}_6$ values, as in Eqs.~\eqref{eq:E6gauge} and \eqref{eq:E6yuk}, while the usual PS matching is applied above $m_\text{KK}$. 
This plot clearly demonstrates that the gauge and Yukawa couplings of the third generation do unify to a single value thanks to the UV fixed point, independently on the value of $m_\text{KK}$.  However, due to the constrained bulk structure, the light generations must be localised on one of the two boundaries.
Before addressing this issue, there are two related features of the bulk interactions: baryon number conservation and the cancellation of 4D gauge anomalies.

Regarding the former, we recall that the theory, at the level of the SM gauge invariance, features five $\text{U}(1)$ symmetries. Besides the three gauged ones, $\text{U}(1)_\text{B-L} \supset \text{SU}(4)$, $\text{U}(1)_\text{R} \supset \text{SU}(2)_\text{R}$ and $\text{U}(1)_\psi$, there are two global charges associated to the matter fields, $\text{U}(1)_{27}$ and $\text{U}(1)_{27'}$. Among the first three, the SM hypercharge is defined as usual in PS models as $2 Q_\text{Y} = Q_\text{B-L} + Q_\text{R}$. There is a single global charge that remains after the breaking of PS and $\text{U}(1)_\psi$, and that is not carried by the Higgs fields:
\begin{equation} \label{eq:baryonN}
Q_\text{B} = \frac{1}{4} Q_\text{B-L} + \frac{1}{6} Q_{27} + \frac{1}{12} Q_\psi - \frac{1}{3} Q_{27'} \,,
\end{equation}
where the global charges are normalised to unity. On the SM fields, this charge matches baryon number. Being respected by all bulk interactions, it protects the proton from decaying. 
As in the minimal $\text{SU}(5)$ aGUT model \cite{Cacciapaglia:2020qky}, the components of the bulk fields with opposite parities on the two boundaries, shown by the black symbols in Fig.~\ref{fig:irreps}, have unusual $Q_\text{B}$ assignments, hence the lightest state among them is stable. All zero modes have standard baryon number assignments, including the non-SM states. A detailed description of the  components is presented in the supplementary material.

Another feature of the bulk model is the presence of 4D anomalies at the level of the gauge symmetry PS$\times \text{U}(1)_\psi$, which can only stem from the matter fields as the ${\bf 78}$ is real. As shown in Fig.~\ref{fig:irreps}, however, the zero modes of the ${\bf 27}$ and ${\bf 27'}$ form effectively complete representations of $\text{SO}(10) \supset\text{PS}$, namely $16_1 + 10_2 + 1_{-4}$. Hence, the PS gauge symmetry is anomaly-free. For instance, the $16_1$ is formed by the $({\bf 4}, {\bf 2}, {\bf 1})_1$ zero mode in $\Phi_{27}$ and the $(\bar{\bf 4}, {\bf 1}, {\bf 2})_1$ in $\Phi_{27'}^c$. The 4D anomalies, therefore, only involve the $\text{U}(1)_\psi$ charges, where the coefficients respect $\mathcal{A}(16_1) = \mathcal{A} (10_2 + 1_{-4})$. The $\text{U}(1)_\psi$ gauge anomalies can only be cancelled by adding fields on the $\text{SO}(10)$ boundary. This leads to two possible models:
\begin{itemize}
    \item[i)] Two $16_{-1}$ multiplets, whose components match the two SM light generations. Hence, the total number of SM generations is \emph{predicted to be three} by gauge anomaly cancellation.  However, baryon number is violated by the localised gauge and Yukawa interactions. 
    \item[ii)] Two $10_{-2} + 1_{4}$, corresponding to massive states, while the light generations are localised on the other boundary.
\end{itemize}
In the first case i), the two localised superfields $\Phi^i_{16_{-1}}$ allow for the following couplings to the bulk field containing the Higgs doublets:
\begin{equation}
    y^{ij}\ \Phi^i_{16_{-1}} \Phi^j_{16_{-1}} \left. \Phi^c_{27}\right|_{10_2}\,.
\end{equation}
This term, and the $\text{SO}(10)$ gauge interactions of the localised fields, violate baryon and lepton numbers, as in traditional GUT models, via couplings with the first KK resonances. For flat extra dimensions, this would lead to a direct bound $m_\text{KK} \gtrsim 10^{16}$~GeV \cite{Super-Kamiokande:2009yit}.  
We remark that warping the extra space can lead to a mild suppression of these coupling \cite{Pomarol:1999ad}, hence it would be feasible to lower the KK scale by one or two orders of magnitude. Furthermore, additional localised superfields can be added in order to explain the different values of the Yukawa couplings, as done in minimal $\text{SO}(10)$ GUTs \cite{Aulakh:2003kg}: the presence of large representations does not spoil the UV fixed point as they only contribute to logarithmic running, which is overpowered by the bulk power-law running. As an example, in Fig.~\ref{fig:running} we show a case with $m_\text{KK} = 10^{15}$~GeV. The mixing between the third generation and the light ones, however, is forbidden by the $\text{U}(1)_\psi$ gauge symmetry, and it requires additional localised fields, see supplementary material. 

In the second case ii), one can add two copies of the following set of superfields on the $\text{SU}(6)_\text{L}\times\text{SU}(2)_\text{R}$ boundary:
\begin{equation}
    \Phi^j_{(15,1)}\,, \; \Phi^j_{(\bar{6},2)}\,, \; \Phi^j_{(20,2)}\,, \;\; j=1,2\,.
\end{equation}
The above fields match the bulk field components containing the third generation, hence the three generations share similar localised couplings. The localised Yukawas are written in terms of the following $3\times 3$ matrices
\begin{equation}
    y^{ij}\ \Phi^i_{(15,1)} \Phi^i_{(20,2)} \left. \Phi_{27}^c \right|_{(6,2)} + \lambda^{ij}\ \Phi^i_{(\bar{6},2)} \Phi^j_{(20,2)} \left. \Phi_{27'}^c \right|_{(\bar{15},1)} \,,
\end{equation}
where $i=3$ corresponds to the bulk fields and the second term gives mass to the unwanted components. As the above couplings have the same structure of the bulk ones, the baryon number in Eq.~\eqref{eq:baryonN} remains conserved and the $m_\text{KK}$ scale can be lowered compared to case i). In this model, it can be as low as the lowest scale allowed by PS to be around $2000$~TeV, which is obtained by using the current limit on $\rm{Br}\left(K_L\rightarrow\mu^{\pm}e^{\mp}\right)<4.7\times 10^{-12}$\cite{BNL:1998apv}(see also e.g.~\cite{Volkas:1995yn,Parida:2014dba,Molinaro:2018kjz}).

\vspace{0.3cm}

In conclusion, in this letter we presented a new aGUT  based on a supersymmetric exceptional ${\rm E}_6$ gauge symmetry. Its uniqueness stands in the presence of a single UV fixed point for gauge and Yukawa couplings of the third generation. The number of SM generations is predicted by gauge anomaly cancellation. We highlighted a second option, which preserves baryon number and allows to lower the compactification scale down to a few thousand TeV. This exceptional model has far reaching implications both for low energy phenomenology, for instance in the flavour sector, and at high energies, via new model building opportunities for UV completions.

\textit{Acknowledgements} -- We acknowledge useful suggestions on the ${\rm E}_6$ contractions from B.~Baj{\'c}. R.P.~is supported in part by the Swedish Research Council grant, contract number 2016-05996, as well as by the European Research Council (ERC) under the European Union's Horizon 2020 research and innovation programme (grant agreement No. 668679). Z.W.W.~acknowledges ``Hundred Talents Program" supported by UESTC.

\bibliography{biblio}

\newpage

\widetext
\appendix

\section{Supplementary material}

\subsection{5D supersymmetric Lagrangian}

The gauge fields consist of one 5D vector $V^M$ ($M=1,\dots 5$), one 5D spinor $(\lambda, \lambda')$ and a real scalar $\sigma$, transforming as the adjoint irreducible representation (irrep) of $\text{E}_6$. They can be expressed in terms of the following $\mathcal{N}=1$ superfields:
\begin{equation}
    W^\alpha_{78} = \{ V^\mu,\ \lambda \}\,, \quad \Phi_{78} = \{  \Sigma,\ \lambda'\}\,,
\end{equation}
where $\Sigma = (\sigma + i V^5)$. Here, $W^\alpha$ is a vector superfield, while $\Phi$ is a chiral one. Together, they form a 5D hypermultiplet, which resembles that of $\mathcal{N}=2$ supersymmetry in 4D. The 5D action can be written as:
\begin{equation}
    S_{5D} = \frac{1}{4kg^2} \ \mbox{Tr}\; \left[ \left( \frac{1}{4} W^\alpha W_\alpha\ \delta^2 (\bar \theta) + \mbox{h.c.} \right) + \left( e^{-2gV} \nabla_y e^{2gV}\right)^2 \right]\,,
\end{equation}
where
\begin{eqnarray}
    W_\alpha &=& -\frac{1}{4} \bar{D}^2 e^{-2gV} D_\alpha e^{2gV}\,, \\
    \nabla_y e^{2gV} &=& \partial_y e^{2gV} - g\bar{\Phi} e^{2gV} - e^{2gV} g\Phi\,.
\end{eqnarray}

A 5D matter multiplet in the irrep $R$ consists of a Dirac spinor $\psi_R$ and two scalar states $\phi_{R1}$ and $\phi_{R2}$. Writing the Dirac spinor in terms of 4D Weyl spinors as
\begin{equation}
    \psi_R = \begin{pmatrix} \chi_R \\ \bar{\xi}_R \end{pmatrix}\,,
\end{equation}
the fields can be organised into two chiral superfields
\begin{equation}
    \Phi_R = \{ \phi_{R1},\ \chi_R\} \,, \qquad  \Phi_R^c = \{ \phi^\dagger_{R2},\ \xi_R\} \,, 
\end{equation}
where the second one transforms as conjugate irrep of $\text{E}_6$. In other words, $\Phi$ contains the left-handed spinor, while $\Phi^c$ contains the charge-conjugate right-handed one.
The 5D action, including interactions with the gauge fields, can be written as
\begin{equation}
    S_{5D} = \bar{\Phi}_R e^{2gV} \Phi_R + \bar{\Phi}_R^c e^{-2gV} \Phi_R^c + \left(\Phi_R^c (\nabla_y + M) \Phi \ \delta(\bar{\theta})+ \mbox{h.c.} \right)\,.
\end{equation}
The third term above takes care of the 5D Lorentz invariance by introducing a coupling between the two chiral supermultiplets. Also, the above Lagrangian has a global U(1) symmetry under which $\Phi_R \to e^{i Q_R \theta}\ \Phi_R$ and $\Phi_R^c \to e^{-i Q_R \theta} \Phi_R^c$\,.

\subsection{$\text{E}_6$ breaking and orbifold boundary conditions}

In general, the components of an $\text{E}_6$ multiplet will receive different parity assignments. Let us consider a parity $P$ that breaks $\text{E}_6 \to \mathcal{H}$, so that an irrep $R$ of $\text{E}_6$ decomposes in components $\chi_j$. The parity assignments of each component $\chi_j$ will read:
\begin{equation}
    P (\chi_j) = \eta_R \ \mathcal{P}_j\,,
\end{equation}
where $\eta_R$ is an overall sign that can be chosen different for different fields in the irrep $R$, while the relative parities $\mathcal{P}_j$ only depend on the $\text{E}_6$ irrep (and are the same for all fields).

We consider the following breaking patterns:
\begin{eqnarray}
\text{A}:\; \text{E}_6 &\to& \text{SO}(10) \times \text{U}(1)_\psi\,, \\
\text{B}:\; \text{E}_6 &\to& \text{SU}(6)_\text{L} \times \text{SU}(2)_\text{R}\,, \\
\text{C}:\; \text{E}_6 &\to& \text{SU}(6)_\text{R} \times \text{SU}(2)_\text{L}\,.
\end{eqnarray}
The intersection of any pair of them leaves the Pati-Salam group unbroken, $\text{SU}(4)\times \text{SU}(2)_\text{L} \times \text{SU}(2)_\text{R}$, plus $\text{U}(1)_\psi$. 

For the fundamental, ${\bf 27}$, we have the following decompositions:
\begin{eqnarray}
\begin{split}
\text{E}_6\rightarrow \text{SO}(10)_{Q_\psi} &\rightarrow (\text{SU}(4)\times \text{SU}(2)_\text{L} \times \text{SU}(2)_\text{R})_{Q_\psi} \\
(16)_{1} \,&=\, \left(4,2,1\right)_1\, + \,\left(\bar{4},1,2\right)_1 \\
(10)_{-2} \,&=\, \left(6,1,1\right)_{-2}\, + \,\left(1,2,2\right)_{-2}\\
(1)_{4}\,&=\, \left(1,1,1\right)_4\,;
\end{split}
\end{eqnarray}
\begin{eqnarray}
\begin{split}
\text{E}_6\rightarrow \text{SU}(6)_\text{L}\times\text{SU}(2)_\text{R} &\rightarrow (\text{SU}(4)\times \text{SU}(2)_\text{L} \times \text{SU}(2)_\text{R})_{Q_\psi} \\
\left(\bar{6},2\right) \,&=\, \left(\bar{4},1,2\right)_1\, + \,\left(1,2,2\right)_{-2} \\
\left(15,1\right) \,&=\, \left(4,2,1\right)_1\, + \,\left(6,1,1\right)_{-2}\, +\,\left(1,1,1\right)_4\,;
\end{split}
\end{eqnarray}
\begin{eqnarray}
\begin{split}
\text{E}_6\rightarrow \text{SU}(6)_\text{R}\times\text{SU}(2)_\text{L} &\rightarrow (\text{SU}(4)\times \text{SU}(2)_\text{L} \times \text{SU}(2)_\text{R})_{Q_\psi}\\
\left(6,2\right) \,&=\, \left(4,2,1\right)_1\, + \,\left(1,2,2\right)_{-2} \\
\left(\bar{15},1\right) \,&=\, \left(\bar{4},1,2\right)_1\, + \,\left(6,1,1\right)_{-2}\, +\,\left(1,1,1\right)_4\,.
\end{split}
\end{eqnarray}
The internal parities $\mathcal{P}_j$ are listed in Table~\ref{tab:27}.

\begin{table}[htb]
\centering
\begin{tabular}{l|c|c|c|}
\phantom{\big(}$\bf 27$ & $\text{SO}(10)\times\text{U}(1)_\psi$ & $\text{SU}(6)_\text{L}\times \text{SU}(2)_\text{R}$ & $\text{SU}(6)_\text{R}\times \text{SU}(2)_\text{L}$ \\ \hline\hline
\phantom{\big(}$(4,2,1)_1$ & \textcolor{blue}{even} & \textcolor{blue}{even} & \textcolor{red}{odd} \\ \hline
\phantom{\big(}$(\bar{4},1,2)_1$ & \textcolor{blue}{even} & \textcolor{red}{odd} & \textcolor{blue}{even} \\ \hline
\phantom{\big(}$(6,1,1)_{-2}$ & \textcolor{red}{odd} & \textcolor{blue}{even} & \textcolor{blue}{even} \\ \hline 
\phantom{\big(}$(1,2,2)_{-2}$ & \textcolor{red}{odd} & \textcolor{red}{odd} & \textcolor{red}{odd} \\ \hline
\phantom{\big(}$(1,1,1)_4$ & \textcolor{green}{odd} & \textcolor{blue}{even} & \textcolor{blue}{even} \\ \hline
\end{tabular}
\caption{\label{tab:27} Intrinsic parities for the fundamental $\bf 27$ of $\text{E}_6$. The colours label PS fields in the same irrep of $\mathcal{H}$.}
\end{table}

For the adjoint, $\bf 78$, we have the following decompositions:
\begin{eqnarray}
\begin{split}
\text{E}_6\rightarrow \text{SO}(10)_{Q_\psi} &\rightarrow (\text{SU}(4)\times \text{SU}(2)_\text{L} \times \text{SU}(2)_\text{R})_{Q_\psi} \\
(45)_{0} \,&=\, \left(15,1,1\right)_0\,+\,\left(1,3,1\right)_0\,+\,\left(1,1,3\right)_0\,+\,\left(6,2,2\right)_0 \\
(16)_{-3} \,&=\, \left(4,2,1\right)_{-3}\, + \,\left(\bar{4},1,2\right)_{-3} \\
(\bar{16})_{3} \,&=\, \left(\bar{4},2,1\right)_3\, + \,\left(4,1,2\right)_3 \\
(1)_{0}\,&=\, \left(1,1,1\right)_0\,;
\end{split}
\end{eqnarray}
\begin{eqnarray}
\begin{split}
\text{E}_6\rightarrow \text{SU}(6)_\text{L}\times\text{SU}(2)_\text{R} &\rightarrow (\text{SU}(4)\times \text{SU}(2)_\text{L} \times \text{SU}(2)_\text{R})_{Q_\psi} \\
\left(35,1\right) \,&=\, \left(15,1,1\right)_0\,+\, \left(1,3,1\right)_0\,+\,\left(4,2,1\right)_{-3}\,+\,\left(\bar{4},2,1\right)_{3}\,+\, \left(1,1,1\right)_0 \\
\left(1,3\right) \,&=\, \left(1,1,3\right)_0 \\
\left(20,2\right) \,&=\, \left(6,2,2\right)_0\, + \,\left(\bar{4},1,2\right)_{-3}\,+\,\left(4,1,2\right)_{3}\,;
\end{split}
\end{eqnarray}
\begin{eqnarray}
\begin{split}
\text{E}_6\rightarrow \text{SU}(6)_\text{R}\times\text{SU}(2)_\text{L} &\rightarrow (\text{SU}(4)\times \text{SU}(2)_\text{L} \times \text{SU}(2)_\text{R})_{Q_\psi} \\
\left(35,1\right) \,&=\, \left(15,1,1\right)_0\,+\, \left(1,1,3\right)_0\,+\,\left(4,1,2\right)_3\,+\,\left(\bar{4},1,2\right)_{-3}\,+\, \left(1,1,1\right)_0 \\
\left(3,1\right) \,&=\, \left(1,3,1\right)_0 \\
\left(20,2\right) \,&=\, \left(6,2,2\right)_0\, + \,\left(\bar{4},2,1\right)_3\,+\,\left(4,2,1\right)_{-3}\,.
\end{split}
\end{eqnarray}

The internal parities $\mathcal{P}_j$ are listed in Table~\ref{tab:adj}. Note that for the gauge fields, the overall signs are fixed to be $\eta=+1$ for $W^\alpha$ and $\eta=-1$ for $\Phi$ for all parities. Hence, for the choice of parities B--C, it is not possible to obtain zero modes in the fourplets of $\text{SU}(4)$ in $\phi$, which could play the role of gaugino matter fields in the SM.
Instead, for the choice A--C, the gauginos contain the left-handed $\text{SU}(2)_\text{L}$ doublets, hence they cannot be used for the breaking of the PS gauge symmetry via the Scherk-Schwarz mechanism.

\begin{table}[htb]
\centering
\begin{tabular}{l|c|c|c|}
\phantom{\big(}$\bf 78$ & $\text{SO}(10)\times\text{U}(1)_\psi$ & $\text{SU}(6)_\text{L}\times \text{SU}(2)_\text{R}$ & $\text{SU}(6)_\text{R}\times \text{SU}(2)_\text{L}$ \\ \hline\hline
\phantom{\big(}$(15,1,1)_0$ & \textcolor{blue}{even} & \textcolor{blue}{even} & \textcolor{blue}{even}\\ \hline
\phantom{\big(}$(1,3,1)_0$ & \textcolor{blue}{even} & \textcolor{blue}{even} & \textcolor{green}{even}\\ \hline
\phantom{\big(}$(1,1,3)_0$ & \textcolor{blue}{even} & \textcolor{green}{even} & \textcolor{blue}{even}\\ \hline
\phantom{\big(}$(6,2,2)_0$ & \textcolor{blue}{even} & \textcolor{red}{odd} & \textcolor{red}{odd}\\ \hline 
\phantom{\big(}$(4,2,1)_{-3}$ & \textcolor{red}{odd} & \textcolor{blue}{even} & \textcolor{red}{odd}\\ \hline
\phantom{\big(}$(\bar{4},1,2)_{-3}$ & \textcolor{red}{odd} & \textcolor{red}{odd} & \textcolor{blue}{even}\\ \hline
\phantom{\big(}$(\bar{4},2,1)_3$ & \textcolor{green}{odd} & \textcolor{blue}{even} & \textcolor{red}{odd}\\ \hline
\phantom{\big(}$(4,1,2)_3$ & \textcolor{green}{odd} & \textcolor{red}{odd} & \textcolor{blue}{even}\\ \hline
\phantom{\big(}$(1,1,1)_0$ & even & \textcolor{blue}{even} & \textcolor{blue}{even}\\ \hline
\end{tabular}
\caption{\label{tab:adj} Intrinsic parities for the adjoint $\bf 78$ of $\text{E}_6$. The colours label PS fields in the same irrep of $\mathcal{H}$.}
\end{table}

A detailed list of the bulk field components with their assigned orbifold parities can be found in Tables~\ref{tab:5Dmatter}, \ref{tab:5DmatterX} and \ref{tab:5Dgauge}, for the choice A--B, which is presented in the main text.

\subsection{Baryon number is conserved}

At the level of the SM gauge symmetry, the model has 5 U(1) factors: $\text{U}(1)_\text{B-L}$ in $\text{SU}(4)$, $\text{U}(1)_\text{R}$ in $\text{SU}(2)_\text{R}$, $\text{U}(1)_\psi$, $\text{U}(1)_{27}$ and $\text{U}(1)_{27'}$. The two last ones are global, while the first 3 are gauged.

Two combinations are broken by the vacuum expectation values that break PS$\times U(1)_\psi$ down to the SM:
\begin{eqnarray}
    \langle ({\bf 1},{\bf 1},{\bf 1})_{-4} \rangle\;\; \mbox{in}\;\; {\bf 27'} &\Rightarrow& -4 Q_\psi - Q_{27'} = 0 \\
     \langle ({\bf 4},{\bf 1},{\bf 2})_{3} \rangle\;\; \mbox{in}\;\; {\bf 78} &\Rightarrow& Q_\text{R}-Q_\text{B-L}+3 Q_\psi= 0
\end{eqnarray}
where we assume that the PS breaking is due to the neutrino singlets zero modes contained in the adjoint (via the Scherk-Scharz mechanism). 
We can further define the hypercharge as follows:
\begin{equation}
    Q_\text{Y} = \frac{1}{2} (Q_\text{B-L} + Q_\text{R})\,.
\end{equation}
Henceforth, it remains the freedom of defining two unbroken global symmetries. There is a unique charge which vanishes on both Higgs doublets, hence it will not be broken by the Higgs vacuum expectation values:
\begin{equation}
Q_\text{B} = \frac{1}{4} Q_\text{B-L} + \frac{1}{6} Q_{27} + \frac{1}{12} Q_\psi - \frac{1}{3} Q_{27'} \,,
\end{equation}
where the normalisation matches the baryon number of quarks.
The second combination must be non-zero on at least one of the Higgs doublets: choosing to normalise the charge to be 1 on $\varphi_{h2}$, we obtain 
\begin{equation}
Q_\text{K} = \frac{1}{2} Q_\text{R}  + \frac{1}{6} Q_{27} - \frac{1}{6} Q_\psi + \frac{2}{3} Q_{27'} \,.
\end{equation}
The charge assignments are listed in Tables~\ref{tab:5Dmatter}, \ref{tab:5DmatterX} and~\ref{tab:5Dgauge}.

Note that the PS symmetry may also be broken by the $(\bar{\bf 4},{\bf 1},{\bf 2})_1$ in ${\bf 27'}$, however this leads to the same charge definition as above.

\subsection{Localised superpotential on the SO(10) boundary}

In the model with two localised generations on the SO$(10)$ boundary, mixings with the bulk third generation are not possible, as they are forbidden by the $Q_\psi$ charges. As a solution, one can introduce a minimal set of localised fields, such as a set of localised Higgs states in the $10_0$, which allows for
\begin{equation}
    \tilde{y}^i\ \Phi^i_{16_{-1}} \left. \Phi_{27}\right|_{16_1} \Phi_{10_0}\,,
\end{equation}
hence generating a mixing in the left-handed sector of the theory. Note that no additional gauge anomalies are generated by such localised Higgs states. A mixing between the localised and bulk Higgs fields could be generated by SUSY-breaking quartic couplings, or by introducing additional singlets charged under U$(1)_\psi$ yielding the following superpotential:
\begin{equation}
    a\ \Phi_{1_2} \Phi_{1_{-2}} + b\ \Phi_{10_0} \left. \Phi_{27}^c \right|_{10_2} \Phi_{1_{-2}} + c\ \Phi_{1_{2}} \Phi_{1_2} \left. \Phi_{27'}^c \right|_{1_{-4}}\,.
\end{equation}

\newpage

\begin{table}[htb]
\centering\begin{tabular}{|c|c|c|c|c|c|c||c|c|c|c|}
\hline
 $\text{E}_6$ fields &  PS fields & $(\mathbb{Z}_2,\mathbb{Z}_2')$ & SM & name &$Q_\text{B}$&$Q_\text{K}$&$Q_\text{R}$&$Q_\text{B-L}$&$Q_\psi$&$Q_{27}$ \\
\hline
$\Phi_{27}$ & $(\bf{4}, \bf{2}, \bf{1})_{1}$ & $(+,+)$ & $(3,2)_{1/6}$ & $q_L$ & $1/3$ & $0$ &$0$&$1/3$&$1$&$1$ \\
& & & $(1,2)_{-1/2}$ & $l_L$ & $0$ & $0$ &$0$&$-1$&$1$&$1$ \\ \cline{2-7}
& $(\bar{\bf 4}, {\bf 1}, {\bf 2})_{1}$ & $(+,-)$ & $(\bar{3},1)_{1/3}$ & & $1/6$ & $1/2$ &$1$&$-1/3$&$1$&$1$ \\
&&& $(\bar{3},1)_{-2/3}$ & & $1/6$ & $-1/2$ &$-1$&$-1/3$&$1$&$1$  \\
&&& $(1,1)_{1}$ & & $1/2$ & $1/2$ &$1$&$1$&$1$&$1$ \\
&&& $(1,1)_{0}$ & & $1/2$ & $-1/2$  &$-1$&$1$&$1$&$1$ \\\cline{2-7}
& $(\bf{6}, \bf{1}, \bf{1} )_{-2}$ & $(-,+)$ & $(\bar{3},1)_{1/3}$ & & $1/6$ & $1/2$ &$0$&$2/3$&$-2$&$1$ \\
&&& $(3,1)_{-1/3}$ & & $-1/6$ & $1/2$ &$0$&$-2/3$&$-2$&$1$ \\ \cline{2-7}
& $(\bf{1}, \bf{2}, \bf{2} )_{-2}$ & $(-,-)$ & $(1,2)_{1/2}$ & & $0$ & $1$ &$1$&$0$&$-2$&$1$ \\
&&& $(1,2)_{-1/2}$ & & $0$ & $0$ &$-1$&$0$&$-2$&$1$ \\ \cline{2-7}
& $(\bf{1},\bf{1},\bf{1})_{4}$ & $(-,+)$ & $(1,1)_0$ & & $1/2$ & $-1/2$ &$0$&$0$&$4$&$1$ \\
\hline
$\Phi^c_{27}$ & $(\bar{\bf{4}}, \bf{2}, \bf{1})_{-1}$ & $(-,-)$ & $(\bar{3},1)_{-1/6}$ &  & $-1/3$ & $0$ &$0$&$-1/3$&$-1$&$-1$ \\
& & & $(1,2)_{1/2}$ &  & $0$ & $0$ &$0$&$1$&$-1$&$-1$ \\ \cline{2-7}
& $(\bf 4, {\bf 1}, {\bf 2})_{-1}$ & $(-,+)$ & $(3,1)_{-1/3}$ & & $-1/6$ & $-1/2$ &$-1$&$1/3$&$-1$&$-1$ \\
&&& $(3,1)_{2/3}$ & & $-1/6$ & $1/2$ &$1$&$-1/3$&$-1$&$-1$ \\
&&& $(1,1)_{-1}$ & & $-1/2$ & $-1/2$ &$-1$&$-1$&$-1$&$-1$ \\
&&& $(1,1)_{0}$ & & $-1/2$ & $1/2$ &$1$&$-1$&$-1$&$-1$ \\\cline{2-7}
& $(\bf{6}, \bf{1}, \bf{1} )_{2}$ & $(+,-)$ & $(\bar{3},1)_{1/3}$ & & $1/6$ & $-1/2$ &$0$&$2/3$&$2$&$-1$ \\
&&& $(3,1)_{-1/3}$ & & $-1/6$ & $-1/2$ &$0$&$-2/3$&$2$&$-1$ \\ \cline{2-7}
& $(\bf{1}, \bf{2}, \bf{2} )_{2}$ & $(+,+)$ & $(1,2)_{1/2}$ & $\varphi_{h1}$ & $0$ & $0$ &$1$&$0$&$2$&$-1$ \\
&&& $(1,2)_{-1/2}$ & $\varphi_{h2}^\ast$ & $0$ & $-1$ &$-1$&$0$&$2$&$-2$ \\ \cline{2-7}
& $(\bf{1},\bf{1},\bf{1})_{-4}$ & $(+,-)$ & $(1,1)_0$ & & $-1/2$ & $1/2$ &$0$&$0$&$-4$&$-1$ \\
\hline
\end{tabular}
\caption{\label{tab:5Dmatter} Field decomposition for the Matter multiplet $R={\bf 27}$.}
\end{table}

\begin{table}[htb]
\centering\begin{tabular}{|c|c|c|c|c|c|c||c|c|c|c|}
\hline
 $\text{E}_6$ fields &  PS fields & $(\mathbb{Z}_2,\mathbb{Z}_2')$ & SM & name & $Q_\text{B}$&$Q_\text{K}$ & $Q_\text{R}$ & $Q_\text{B-L}$ & $Q_{\psi}$ & $Q_{27'}$ \\
\hline
$\Phi_{27'}$ & $(\bf{4}, \bf{2}, \bf{1})_{1}$ & $(+,-)$ & $(3,2)_{1/6}$ &  & $-1/6$ & $1/2$ &$0$& $1/3$&$1$&$1$\\
& & & $(1,2)_{-1/2}$ &  & $-1/2$ & $-1/2$ &$0$&$-1$&$1$&$1$\\ \cline{2-7}
& $(\bar{\bf 4}, {\bf 1}, {\bf 2})_{1}$ & $(+,+)$ & $(\bar{3},1)_{1/3}$ & ${b'}_R^c$ & $-1/3$ & $1$ &$1$&$-1/3$&$1$&$1$ \\
&&& $(\bar{3},1)_{-2/3}$ & ${t'}_R^c$ & $-1/3$ 
& $0$ &$-1$&$-1/3$&$1$&$1$  \\
&&& $(1,1)_{1}$ & ${\tau'}_R^c$ & $0$ & $1$ &$1$&$1$&$1$&$1$ \\
&&& $(1,1)_{0}$ & ${\nu'}_R^c$ & $0$ & $0$  &$-1$&$1$&$1$&$1$ \\\cline{2-7}
& $(\bf{6}, \bf{1}, \bf{1} )_{-2}$ & $(-,-)$ & $(\bar{3},1)_{1/3}$ & & $-1/3$ & $1$ &$0$&$2/3$&$-2$&$1$ \\
&&& $(3,1)_{-1/3}$ & & $1/3$ & $1$ &$0$&$-2/3$&$-2$&$1$ \\ \cline{2-7}
& $(\bf{1}, \bf{2}, \bf{2} )_{-2}$ & $(-,+)$ & $(1,2)_{1/2}$ & & $-1/2$ & $3/2$ &$1$&$0$&$-2$&$1$ \\
&&& $(1,2)_{-1/2}$ & & $-1/2$ & $1/2$ &$-1$&$0$&$-2$&$1$ \\ \cline{2-7}
& $(\bf{1},\bf{1},\bf{1})_{4}$ & $(-,-)$ & $(1,1)_0$ & & $0$ & $0$ &$0$&$0$&$4$&$1$ \\
\hline
$\Phi^c_{27'}$ & $(\bar{\bf{4}}, \bf{2}, \bf{1})_{-1}$ & $(-,+)$ & $(\bar{3},2)_{-1/6}$ &  & $1/6$ & $-1/2$ &$0$&$-1/3$&$-1$&$-1$ \\
& & & $(1,2)_{1/2}$ &  & $1/2$ & $1/2$ &$0$&$1$&$-1$&$-1$ \\ \cline{2-7}
& $(\bf 4, {\bf 1}, {\bf 2})_{-1}$ & $(-,-)$ & $(3,1)_{-1/3}$ & & $1/3$ & $-1$ &$-1$&$1/3$&$-1$&$-1$ \\
&&& $(3,1)_{2/3}$ & & $1/3$ & $0$ &$1$&$1/3$&$-1$&$-1$ \\
&&& $(1,1)_{-1}$ & & $0$ & $-1$ &$-1$&$-1$&$-1$&$-1$ \\
&&& $(1,1)_{0}$ & & $0$ & $0$ &$1$&$-1$&$-1$&$-1$ \\\cline{2-7}
& $(\bf{6}, \bf{1}, \bf{1} )_{2}$ & $(+,+)$ & $(\bar{3},1)_{1/3}$ & $B_R^c$ & $-1/3$ & $-1$ &$0$&$2/3$&$2$&$-1$ \\
&&& $(3,1)_{-1/3}$ & $B_R$ & $1/3$ & $-1$ &$0$&$-2/3$&$2$&$-1$ \\ \cline{2-7}
& $(\bf{1}, \bf{2}, \bf{2} )_{2}$ & $(+,-)$ & $(1,2)_{1/2}$ & & $1/2$ & $-1/2$ &$1$&$0$&$2$&$-1$ \\
&&& $(1,2)_{-1/2}$ & & $1/2$ & $-3/2$ &$-1$&$0$&$2$&$-1$ \\ \cline{2-7}
& $(\bf{1},\bf{1},\bf{1})_{-4}$ & $(+,+)$ & $(1,1)_0$ & $\varphi_\psi$ & $0$ & $0$ &$0$&$0$&$-4$&$-1$ \\
\hline
\end{tabular}
\caption{\label{tab:5DmatterX} Field decomposition for the Matter multiplet $R={\bf 27'}$.}
\end{table}

\begin{table}[htb]
\centering\begin{tabular}{|c|c|c|c|c|c|c||c|c|c|}
\hline
 $\text{E}_6$ fields &  PS fields & $(\mathbb{Z}_2,\mathbb{Z}_2')$ & SM & name & $Q_\text{B}$&$Q_\text{K}$ & $Q_\text{R}$ & $Q_\text{B-L}$ & $Q_{\psi}$ \\
\hline
$W^\alpha_{78}$/$\Phi_{78}$ & $(\bf{15}, \bf{1}, \bf{1})_{0}$ & $(+,+)$ & $(8,1)_{0}$ & $G_\mu$ & $0$ & $0$ &$0$& $0$&$0$\\
& & & $(1,1)_{0}$ & $A^{B-L}_\mu$ & $0$ & $0$ &$0$&$0$&$0$ \\
& & & $(3,1)_{2/3}$ & $X_\mu$ & $1/3$ & $0$ &$0$&$4/3$&$0$\\
& & & $(\bar{3},1)_{-2/3}$ & $X^\dagger_\mu$ & $-1/3$ & $0$ &$0$&$-4/3$&$0$\\\cline{2-7}
&$(\bf{1},\bf{3},\bf{1})_0$&$(+,+)$&$(1,3)_0$&$W_\mu$& $0$ & $0$ &$0$&$0$&$0$\\ \cline{2-7}
&$(\bf{1},\bf{1},\bf{3})_0$&$(+,+)$&$(1,1)_1$&$W^{R+}_\mu$& $0$ & $0$ &$2$&$0$&$0$\\ 
&&&$(1,1)_{-1}$&$W_\mu^{R-}$& $0$ & $0$ &$-2$&$0$&$0$\\ 
&&&$(1,1)_{0}$&$A^R_\mu$& $0$ & $0$ &$0$&$0$&$0$\\ \cline{2-7}
&$(\bf{1},\bf{1},\bf{1})_0$&$(+,+)$&$(1,1)_0$&$A^\psi_\mu$& $0$ & $0$ &$0$&$0$&$0$\\ \cline{2-7}
&$(\bf{6},\bf{2},\bf{2})_0$&$(+,-)$&$(\bar{3},2)_{5/6}$&& $1/6$ & $1/2$ &$1$&$2/3$&$0$\\
&&&$(\bar{3},2)_{-1/6}$&& $1/6$ & $-1/2$ &$-1$&$2/3$&$0$\\
&&&$(3,2)_{1/6}$&& $-1/6$ & $1/2$ &$1$&$-2/3$&$0$\\
&&&$(\bar{3},2)_{-5/6}$&& $-1/6$ & $-1/2$ &$-1$&$-2/3$&$0$\\ \cline{2-7}
&$(\bf{4},\bf{2},\bf{1})_{-3}$&$(-,+)$&$(3,2)_{1/6}$& & $-1/6$ & $1/2$ &$0$&$1/3$&$-3$\\
&&&$(1,2)_{-1/2}$& & $-1/2$ & $1/2$ &$0$&$-1$&$-3$\\ \cline{2-7}
&$(\bar{\bf{4}},\bf{2},\bf{1})_{3}$&$(-,+)$&$(\bar{3},2)_{-1/6}$& & $1/6$ & $-1/2$ &$0$&$-1/3$&$3$\\
&&&$(1,2)_{1/2}$& & $1/2$ & $-1/2$ &$0$&$1$&$3$\\ \cline{2-7}
&$(\bar{\bf{4}},\bf{1},\bf{2})_{-3}$&$(-,-)$&$(\bar{3},1)_{1/3}$& $b_R^c$ & $-1/3$ & $1$ &$1$&$-1/3$&$-3$\\
&&&$(\bar{3},1)_{-2/3}$& $t_R^c$ & $-1/3$ & $0$ &$-1$&$-1/3$&$-3$\\
&&&$(1,1)_{1}$& $\tau_R^c$ & $0$ & $1$ &$1$&$1$&$-3$\\
&&&$(1,1)_{0}$& $\nu_R^c$ & $0$ & $0$ &$-1$&$1$&$-3$\\\cline{2-7}
&$(\bf{4},\bf{1},\bf{2})_{3}$&$(-,-)$&$(3,1)_{-1/3}$& $b'_L$ & $1/3$ & $-1$ &$-1$&$1/3$&$3$\\
&&&$(3,1)_{2/3}$& $t'_L$ & $1/3$ & $0$ &$1$&$1/3$&$3$\\
&&&$(1,1)_{-1}$& $\tau'_L$ & $0$ & $-1$ &$-1$&$-1$&$3$\\
&&&$(1,1)_{0}$& $\nu'_L$ & $0$ & $0$ &$1$&$-1$&$3$\\
\hline
\end{tabular}
\caption{\label{tab:5Dgauge} Field decomposition for the gauge multiplet $R={\bf 78}$. 
Fields with orbifold parities $(+,+)$ have a zero mode in $W^\alpha_{78}$, while $(-,-)$ entails a zero mode in $\Phi_{78}$.}
\end{table}

\end{document}